\documentclass[a4paper, 10pt, conference]{ieeeconf}      
\usepackage{amsmath,amssymb}

\usepackage{color,soul}

\IEEEoverridecommandlockouts                              
\usepackage{graphicx}
\graphicspath{ {./images/} }                                                          
\usepackage{amsmath}                                                          

\usepackage{lipsum}
 
\usepackage{algorithm}
\usepackage[noend]{algpseudocode}
\usepackage{subcaption}
\DeclareCaptionSubType*{algorithm}

\DeclareCaptionLabelFormat{alglabel}{Alg.~#2}
\usepackage{xcolor}

\usepackage{hyperref}

\usepackage{tabu}

\title{\LARGE \bf
Accelerated  Search for Non\textendash Negative Greedy Sparse Decomposition via Dimensionality Reduction
}

\author{ Konstantinos A. Voulgaris \quad Mike E. Davies \quad Mehrdad Yaghoobi \newline \\ Institute for Digital Communications, the University of Edinburgh, EH9 3JL, UK \\ $\left \{\textrm{Konstantinos.Voulgaris, Mike.Davies, myvaigha}\right \}$@ed.ac.uk }
\begin{document}

\maketitle
\thispagestyle{empty}
\pagestyle{empty}

\begin{abstract}

Non\textendash negative signals form an important class of sparse signals. Many algorithms have already been
proposed to recover such non-negative representations, where
greedy and convex relaxed algorithms are among the most
popular methods. One fast implementation is the FNNOMP algorithm that updates the non\textendash negative coefficients in an iterative manner. Even though FNNOMP is a good approach when working on libraries of small size, the operational time of the algorithm grows significantly when the size of the library is large. This is mainly due to the selection step of the algorithm that relies on matrix vector multiplications. We here introduce the Embedded Nearest Neighbor (E\textendash NN) algorithm which  accelerates the search over large datasets while it is guaranteed to find the most correlated atoms. We then replace the selection step of FNNOMP by E\textendash NN. Furthermore we introduce the Update Nearest Neighbor (U\textendash NN) at the look up table of FNNOMP in order to assure  the non\textendash negativity  criteria of FNNOMP. The results indicate that the proposed methodology can accelerate FNNOMP with a factor 4  on a real dataset of \textit{Raman} Spectra  and with a factor of 22 on a synthetic dataset.   
\newline \newline
\textbf{Index Terms: Matching Pursuit, Orthogonal Matching Pursuit,
Non-negative Sparse Approximations, 
Non-negative Least Square and Spectral Decomposition, Scalable algorithms, Dimensionality Reduction, Linear Embedding, Raman Spectroscopy}

\end{abstract}

\section{INTRODUCTION}

Let the signal of interest be $y \in R^{M}$ and a dictionary of elements $\Phi \in R^{M\times N}$ be given. The linear sparse approximation can be formulated as finding the sparsest $x \in R^{N}$, $M<N$, i.e having the minimum number of non\textendash zero elements, as follows:
\begin{equation}
y\approx \Phi x
\end{equation}
The greedy sparse approximation algorithms are in general characterized by a low computational cost, suitable for real\textendash time and large scale sparse approximations. The  Orthogonal Matching Pursuit (OMP) \cite{c2},\cite{c3} algorithm is  introduced, to find the best representation using selected atoms and approximate the sparse solution of the following problem: 
\begin{equation}\label{eq:ref}
\tilde{x}:= \textrm{argmin}_{x_{s}} ||y-\Phi_sx_s||_2\newline
\end{equation}

There are many applications for which the coefficient vectors are not only sparse, but they are also non\textendash negative. Spectral and multi\textendash spectral unmixing, \cite{c4},\cite{c5},  microarray analysis \cite{c6} and  Raman spectral deconvolution \cite{c7} are a few examples.

The original implementation  of OMP has been modified in order to adopt the algorithm to the non\textendash negativity setting for  coefficients. Essentially the original minimization problem introduced in \eqref{eq:ref} is reformulated by adding a constraint that guarantees the non\textendash negativity of the coefficients and takes the following form:
\begin{equation}
\begin{aligned}
& \tilde{x}:= & & \textrm{argmin}_{x_{s}\geq 0} ||y-\Phi_sx_s||_2 \\& \text{} & & ||x_s||_0\leq j
\end{aligned}
\end{equation}\label{eq:ref_1}
The authors in \cite{c8} introduced the Fast Non\textendash Negative Orthogonal Matching Pursuit algorithm which is a greedy technique based on OMP suitable for real\textendash time applications. Even though the implementation of this strategy may be straightforward when considering a dictionary with a relatively small number of atoms, this is not the case when working with a library that contains a significant number of atoms (i.e thousands). This is mainly due to the selection step of the algorithm which has a computational complexity of $\mathcal{O}(MN)$. Consequently the executional time of the algorithm will scale linearly along with the number of  atoms  in $\Phi$. 

Essentially the selection of the best possible candidate within a normalized dictionary can be represented as the Nearest Neighbor Search (NNS): Given a set of points $P= \{p_1,p_2, \cdots,p_n\}$ in a metric space $X$ with distance function $d$, NNS is to efficiently answer queries for finding the closest point in $P$ to $q \in X$. There exist several data structures proposed to fulfill such task such as the \textit{kd}\textendash\textit{tree} \cite{c9} or the \textit{cover tree} \cite{c10}. Due to the curse of dimensionality, it is  unlikely that there exist a general efficient solution to the exact \textit{k}\textendash NN problem \cite{c11}. Approximate algorithms have been proposed to overcome these impracticalities such as the \textit{locality}\textendash \textit{sensitivity hashing} \cite{c12}. Although these algorithms  do not guarantee the acquisition of the exact nearest neighbor they are fast and scalable.  

Within this paper we introduce  an Embedded Nearest Neighbor (E\textendash NN)  in order to reduce the computational cost of the selection step in the algorithm. Considering the dictionary $\Phi$, this means that we shrink the size of $M$ via dimensionality reduction. Given that the data are typically characterized by an intrinsic dimensionality, we are addressing  E\textendash NN as a practical framework that exploits the benefits of conducting the brute force search on the $K$\textendash dimensional subspace compared to the $M$\textendash dimensional original domain. Since it is expected that a mismatch in between the closest point from one domain to the other will occur, we introduce an update step on the algorithm to compensate the error and eventually acquire the exact NN. In that sense, we are considering  E\textendash NN as a bridge between the approximate to the exact solution for the brute force search.

\section{Linear Embeddings}
In this section we introduce the guidelines for reducing the size of  problem via a Linear Embedding. The standard notion regarding dimensionality reduction is that by having an input signal $y \in R^{M}$, the dimension of the signal is reduced via a linear operator $Q: R^{M}\rightarrow R^{K}$, with $K<M$, that embeds the input signal into the lower dimensional space. The projection of the signal $\hat{y}$ in $R^{K}$ is then computed as follows: $\hat{y}=Q y.$

Linear embedding is a standard approach in many applications where we seek for a low\textendash dimensional representation of data living on a high\textendash dimensional space. There exist different methods to perform the embedding, i.e principal component analysis (PCA) \cite{c13}, random projections \cite{cah} etc. A common characteristic of these embeddings is that the relevant position  between library elements is changed when the points are embedded from $R^M$  to $R^K$. In that sense, given a pair of elements $\phi_i, \phi_c \in R^{M}$ and their representations $\hat{\phi}_i,\hat{\phi}_c \in R^{K}$, we usually have: $d(\phi_i,\phi_c)\neq d(\hat{\phi_i},\hat{\phi_c}) $. For an algorithm that searches for the Nearest Neighbor (NN) of  $y$ in $\Phi$, this may lead to a situation in which  $\textrm{NN}_M\neq \textrm{NN}_K$   where NN is the abbreviation for the Nearest Neighbor and $M,K$ corresponds to the dimensions for each domain. At this section we introduce  the Embedded Nearest Neighbor (E\textendash NN) algorithm that under a specific condition the  search in the lower dimensional space eventually  yields the nearest neighbor  in the original domain. In that sense we are seeking an embedding that yields  a minimum distortion from $R^M\rightarrow R^K$. This aspect of the problem can be addressed in terms of a reformulation of the Constructive Johnson\textendash Lindenstrauss \cite{JL} introduced in \eqref{minmax} where $d(b,t)=||b-t||_2$, where $b,t \in \mathcal{A}\subset R^M$. Let $Q$ distorts the distance for at most $\epsilon_{b,t}$. We then have:

\begin{equation}\label{minmax}
\begin{split}
(1-\epsilon_{bt})d(b,t)\leq d(\hat{b},\hat{t})\leq (1+\epsilon_{bt})d(b,t)\\
d(z,t)-\epsilon_{bt}d(b,t)\leq d(\hat{b},\hat{t})\leq d(b,t)+\epsilon_{bt}d(b,t)\\
d(b,t)-\delta\leq d(\hat{b},\hat{t})  \leq d(b,t)+ \delta
\end{split}
\end{equation}
where,
\begin{equation}\label{Hypothesis_1}
    \delta= \max_{b,t \in \mathcal{A}\subset R^M} \ \epsilon_{bt}d(b,t),  
\end{equation}

\textbf{\textit{Lemma}} 1.$\forall b,t \in \mathcal{A}$ with  a $\delta$ coming from \eqref{Hypothesis_1} and   $\forall y \not\in \mathcal{A} $ with $\max \ \epsilon_{yt}d(y,t)\leq \delta $, the E\textendash NN  introduced in Algorithm 1 guarantees the acquisition of the exact NN.
\begin{algorithm}[b!]
\caption{Embedded NN (E\textendash NN)}\label{eq:ring}
\begin{algorithmic}[1]
\State \textbf{Input}: $\Phi, \hat{\Phi},Q,y$.
\State $\hat{y}=Qy$.
\State Form set S=$\big\{ i : \ d(\hat{\phi}_i,\hat{y})\leq  \  d(y,\textrm{NN}_K )+\delta \big\}, \forall \hat{\phi_i}\in \hat{\Phi}$.
\State return \emph{arg} $min_{i \in S } \ d(y,\Phi_i)$ .
\end{algorithmic}
\end{algorithm}

\begin{proof} Considering three points $y,b,t$ where $d(y,b)\leq d(y,t)$. Then there exist 4 characteristic cases for pairwise distances.
\begin{itemize}
    \item Both distances shrink: $d(\hat{y},\hat{b})\leq d(y,b)$, $d(\hat{y},\hat{t})\leq d(y,t)$. Then by incorporating  \eqref{minmax} :\newline
$d(\hat{y},\hat{b})\leq d(y,b)\leq d(y,t)+\delta$.

\item Both distances stretch. Then from  \eqref{minmax} we have:\newline
$
d(\hat{y},\hat{b})-\delta\leq d(y,b)\leq d(y,t)\newline\Rightarrow
d(\hat{y},\hat{b})\leq d(y,t)+\delta
$
\item $d(\hat{y},\hat{b})$ stretches: $d(y,b)\leq d(\hat{y},\hat{b})$, $d(\hat{y},\hat{t})$ shrinks: $d(y,t)\leq d(\hat{y},\hat{t})+\delta$. Then it follows:

$d(\hat{y},\hat{b})\leq d(y,b)+\delta \leq d(y,t)+\delta$.

\item $d(\hat{y},\hat{b})$ shrinks: $d(\hat{y},\hat{b})\leq d(y,b)$, $d(\hat{y},\hat{t})$ stretches: $d(y,t)\leq d(\hat{y},\hat{t})$. Then: \newline
$d(\hat{y},\hat{b})\leq d(y,b)\leq d(y,t). \rlap{$\qquad \Box$}$
\end{itemize}

\end{proof}
The analysis provided by proof of the \textit{Lemma} simply states  that in cases where $NN_M\neq NN_K$, assuming that  $b=NN_M$ and $t=NN_K$  , then $d(\hat{y},\hat{NN}_M)\leq d(y,NN_K)+\delta$.

The complexity of the E\textendash NN introduced in \textbf{Algorithm 1} varies over steps 2\textendash 4 of the algorithm. At step 2 the input signal $y \in R^M$ is embedded in $R^K$ via the linear operator $Q \in R^{K\times M}$. Hence the complexity of step 2 is $\mathcal{O}(KM)$. At step 3 we conduct a number of $N$ distance computations over $M$\textendash dimensional vectors. The computational cost of the corresponding operations is  $\mathcal{O}(KN)$. Finally, at the last of the the algorithm we perform  a number of $|S|$ distance computations on the original space $R^M$. The computational cost of the step is $\mathcal{O}(|S|M)$.

As it can be derived from the analysis there are two critical parameters to benefit from the brute force search in the lower dimensional space. The intrinsic dimensionality of the dataset expressed by $K$ and the cardinality of $S$ on the update step which depends on $\delta$. 

 Essentially we are seeking for an embedding $Q$:
 \begin{equation}\label{minimization}
     \delta= \min_{Q}\ \max_{i,c} \epsilon_{ic}d(\phi_i,\phi_c), \forall i,c \in \Phi.
 \end{equation}

The most common approach to construct  a dimension reduction is principal component analysis (PCA). A key advantage of PCA is that it is computationally  efficient. The embedding to the $K$\textendash dimensional space is simply performed by taking the $K$ dominant eigenvectors of the data covariance matrix. The main drawback of PCA though is that it distorts pairwise distances arbitrarily. In that sense the distance distortion may be significantly larger from the one pair of points to the other. 

An alternative to PCA is the approach of \textit{random projections}. According to the Johnson\textendash Lindenstrauss lemma , given any point cloud $\Omega$ in $R^M$, there exists an embedding $Q$ of dimension $K=\mathcal{O}(\textrm{log}|\Omega|)$ with minimal distortion of the $\binom{|\Omega|}{2}$
 pairwise distances  between the $|\Omega|$ points. This linear embedding is easy to implement in practice. We simply construct a matrix $Q \in R^{K\times M}$ with elements drawn randomly from a certain probability distribution.
 The authors  in \cite{c15}, introduced a deterministic framework, called \textit{NuMax}, that constructs \textit{linear}, \textit{near}\textendash isometric embeddings for data that live in a high\textendash dimensional space. Given a set of training points $\Phi \in R^{M}$, the authors consider a \textit{secant set} \textit{S}($\Phi$) consisting of all pairwise difference vectors of $\Phi$ that lie on the unit sphere. The problem is formulated as an affine rank minimization problem to construct $Q$  such that the norms of all vectors in \textit{S}($\Phi$) are  preserved up to a distortion parameter.

 We aim to solve the problem introduced  in Equation \eqref{minimization} empirically for    library \textit{Raman} spectra with $M=1507$ and $N=4041$ \cite{data} and a library of  Swiss Roll data \cite{c16} which is a synthetic machine Learning dataset of points  that lie on a 2\textendash D manifold but embedded in $R^{1507}$. We found that  the minimization problem introduced by   the \textit{NuMax} algorithm yields a matrix $Q \in R^{K,M}$ with $K=172$ for \textit{Raman} while for the Swiss Roll case $K=3$.
 Then we construct $Q$ for PCA and \textit{random projections} by setting $K=172$ and $K=3$ accordingly such that we can investigate which   method serves the purpose for  $R^{K}$.

 \begin{figure}[t!] 
\centering
\includegraphics[width=0.45\textwidth]{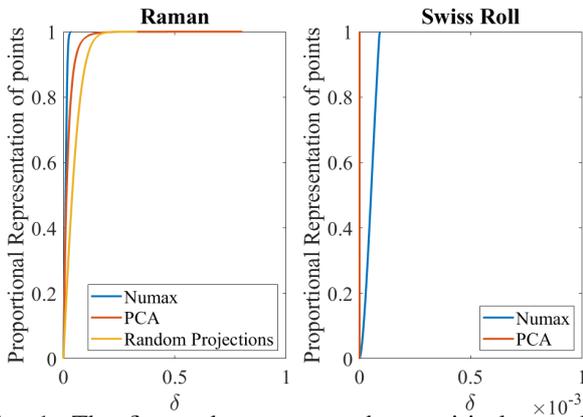}\vspace{-0.3cm} 
\caption{The figure demonstrates the empirical cumulative distribution function (CDF) of $\delta$  over $\Phi$. The distortion on $\Phi$ introduced by random embeddings into Swiss Roll is much larger than \textit{Numax} and PCA hence it is not demonstrated.  }\label{fig:lower}
\end{figure}
 
 The performance for each   method is evaluated with respect  to the error distortion function $\delta(\phi_i,\phi_c)$ as follows:
 \begin{equation}\label{distortion}
 \delta(\phi_i,\phi_c)=|d(\phi_i,\phi_c)-d(\hat{\phi}_i,\hat{\phi}_c)|.
 \end{equation}
 
 The obtained results are demonstrated in figure \ref{fig:lower}.

\subsection{The case of mixtures}
Within our framework we set $\delta$ with respect to the knowledge derived from elements that belong to an available library $\Phi$. The case of mixtures $y$ is slightly different. In particular, each $y$ with sparsity (number of contributing atoms) up to $j$ is formulated as a linear combination of $\phi_i \in \Phi$ as follows: $y=\sum_{w=1}^{j} a_w\phi_w$.

\begin{figure}[t!] 
\centering
\includegraphics[width=0.35\textwidth]{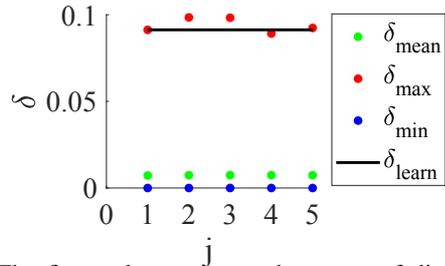}\vspace{-0.3cm} 
\caption{The figure demonstrates the range of distortion over sparsity.$\delta_{mean}(j)=\frac{1}{|\mathcal{Y}||\Phi|}(\sum_{y_m \in \mathcal{Y}}\sum_{\phi_i \in \Phi} \delta(\phi_i,y_m)) ,\newline \delta_{max}(j)=\textrm{max}\ \delta(\phi_i,y_{m}) , \ \delta_{min}(j)=\textrm{min}\ \delta(\phi_i,y_{m})$ }\label{fig:bound}
\end{figure}

 This essentially means that there is not any particular knowledge regarding $\delta (\phi_i,y)$. Hence, an obvious question is whether $y$ is consistent with the choice of $\delta$. Given that according to the results introduced in figure \ref{fig:lower} the $Q$ obtained by the \textit{NuMax} algorithm yields the best results we perform a simulation study for $y$ over a sparsity level up to 5 which is the maximum sparsity of the signals for the  applications we focus on. The distortion is then evaluated according to the error distortion function introduced in \eqref{distortion} with  $y$ taking the place of $\phi_i$ and $\hat{y}$ the place of $\hat{\phi}_i$ accordingly. For each $j$ we generate a set of  mixtures $\mathcal{Y}= \{ y_m \}^L_{m=1}$ via 10000 (denoted as $L$) Monte Carlo simulations.  The obtained results are demonstrated in figure \ref{fig:bound}. Note that $a_w \sim U[0,1]$  and $||y||_2=||\hat{y}||_2=1$. 
 
The results indicate that $\delta$ flunctuates around  $\delta_{learn}$. We empirically observe that the maximum pairwise distortion  $\forall \phi_i \in \Phi$ (denoted as  $\delta_{learn}$)  exceeded only $0.003\%$ over $L$. Even in these cases, the algorithm acquires the exact NN. This is  happening due to the fact that the pairwise distortion  is on average much lower than $\delta_{max}$ and a lower $\delta$ hence  serves the purpose. 

\section{Search FNNOMP}

In this section we introduce an update on the structure of FNNOMP, as introduced in Algorithm 2, with respect to the algorithm introduced in Algorithm 1.  The first change in the structure takes place in the selection step of FNNOMP \cite[pp2]{c7} where we place E\textendash NN. A common phenomenon in sparse non\textendash negative decomposition is that a selected atom may be rejected by the non\textendash negativity criteria introduced in Table I and with respect to equation \eqref{restrict}. Consequently, we need to  modify the content in Table I  compared to the original FNNOMP version.  A key aspect of the changes is the insertion of the U\textendash NN algorithm, as introduced in Algorithm 3, such that E\textendash NN adopts on  the non\textendash negativity setting. All the changes in the overall structure of FNNOMP are highlighted with red. 

In practice Update NN can be addressed as a next NN Algorithm. In that sense anytime that the NN acquired by E\textendash NN  and indexed by $\mu$ is rejected by the criteria introduced

\begin{table}[h]
\centering
\begin{center}
\scalebox{1.24}{\begin{tabular}{|c|c|}
\hline
 if & then \\ \hline
$0< z\leq z^t, z>z^c$ & $z_{j+1}\leftarrow z$, Terminate  \\ \hline
$0< z\leq z^t, z\leq z^c$ &  $z_{j+1}\leftarrow z^c,p\leftarrow p^c$, Terminate  \\ \hline
$z>z^c\geq z^t$ & $p=p+1$,  \textcolor{red}{$\mu$ $\leftarrow $U\textendash NN}  \\ \hline
$z\geq z^c >z^t$  &  $z_{j+1}\leftarrow z^c,p\leftarrow p^c$, Terminate  \\ \hline
$z>z^t>z^c$ & $z^c\leftarrow z^t,p^c\leftarrow p$,  \textcolor{red}{$\mu$ $\leftarrow $U\textendash NN} \\ \hline
$z<0$ & Terminate  \\ \hline
\end{tabular}}
\end{center}\vspace{-0.25cm} \caption{}\label{tab:label}
\end{table}

\begin{equation}\label{restrict}
z_{j+1}\leq z^t= \left\{
\begin{array}{ll}
    \underset{\gamma_i< 0}{\text{min}}  \frac{|x_i|}{|\gamma_i|} & \exists i,\gamma_i\leq 0 \\
     \infty,\         \mathrm{otherwise}  \\
\end{array} 
\right. 
\end{equation}

\begin{algorithm}
\caption{ E\textendash NN on FNNOMP}\label{eq:baseline}
\begin{algorithmic}[1]
\State Initialization: $s=z_0=\emptyset, j=0,r_0=y$.

\State \textbf{while} $j<K$\& max($\Phi^T r_k>0$).
\renewcommand{\labelenumi}{\roman{enumi}}
\begin{enumerate}
\item \textcolor{red}{$\mu\leftarrow$ Embedded\textendash NN.} 
\item $p\leftarrow$ 1.
\item $p^c\leftarrow$ $\mu$.
\item $z^c=0$
\item \textbf{while} $\sim$ Terminate \& $p<N$
\item  \quad \quad $z_t$ from \eqref{restrict}.
\item \quad \quad z$\leftarrow\psi_{\mu}^{T}r_k$: $\psi_{\mu}=\frac{q}{||q||_2}$, $q=(I-\Psi \Psi^{T})\phi_{\mu}$
\item \quad \quad \textcolor{red}{Update based on  Table I}
\item \textbf{end while}
\item $s=s\cup \mu$.
\item Update $\Psi$ and $\textrm{R}^{-1}$
\item $z_{j+1}\leftarrow [z_j,z_{j+1}]$
\item $r_{j+1}=\leftarrow r_j-z_{j+1}\psi_{j+1}$
\item $j\leftarrow j+1$
\end{enumerate}
\State \textbf{end while} .
\State \textbf{output}: $x|_s\leftarrow R^{-1}z_j$
\end{algorithmic}
\end{algorithm}


\begin{algorithm}
\begin{algorithmic}[1]
\State \textbf{Input}: $\Phi, \hat{\Phi},y,\mu,S$.
\State $S=S-\mu$.
\State Form set $S'=\big\{ i : \ d(\hat{\phi}_i,\hat{y})\leq \textrm{min} \ d(y,NN_K )+\delta \big\}$.
\State Form set $S''=S'-S$.
\State return \emph{arg} $min_{\phi_i \in S\cup S'' } \ d(y,S)\cup d(y,S'')$ .
\end{algorithmic}
\caption{\strut Update NN}\label{eq:update}
\end{algorithm}
\vspace{-0.3cm}

in Table II, the task of   U\textendash NN is the acquisition of the next closest point  to $y$.
To do as such we  need  to reject $\mu$  from $S$. This is done in step 2 of the  algorithm.

The implementation of E\textendash NN provides U\textendash NN with the full set of distance measurements in $R^K$ and a number of distance measurements equal to $|S|-1$ in $R^K$ since $\mu$ is rejected in Step 2. Hence no additional distance computation is conducted in Step 3 of U\textendash NN but a simple logical comparison that yields a new set of indexes. Given though that for some of these indexes the distance in $R^M$ is already available from E\textendash NN we introduce Step 4 in order to avoid the recomputation. We then compute the distances for $\phi_i \in S''$ and then we perform a comparison with the measurements of $\phi_i \in S$ in order to find the next NN in $R^M$.

\begin{figure}[t!]
\centering
\begin{subfigure}[b]{0.45\textwidth}
   \includegraphics[width=1\linewidth]{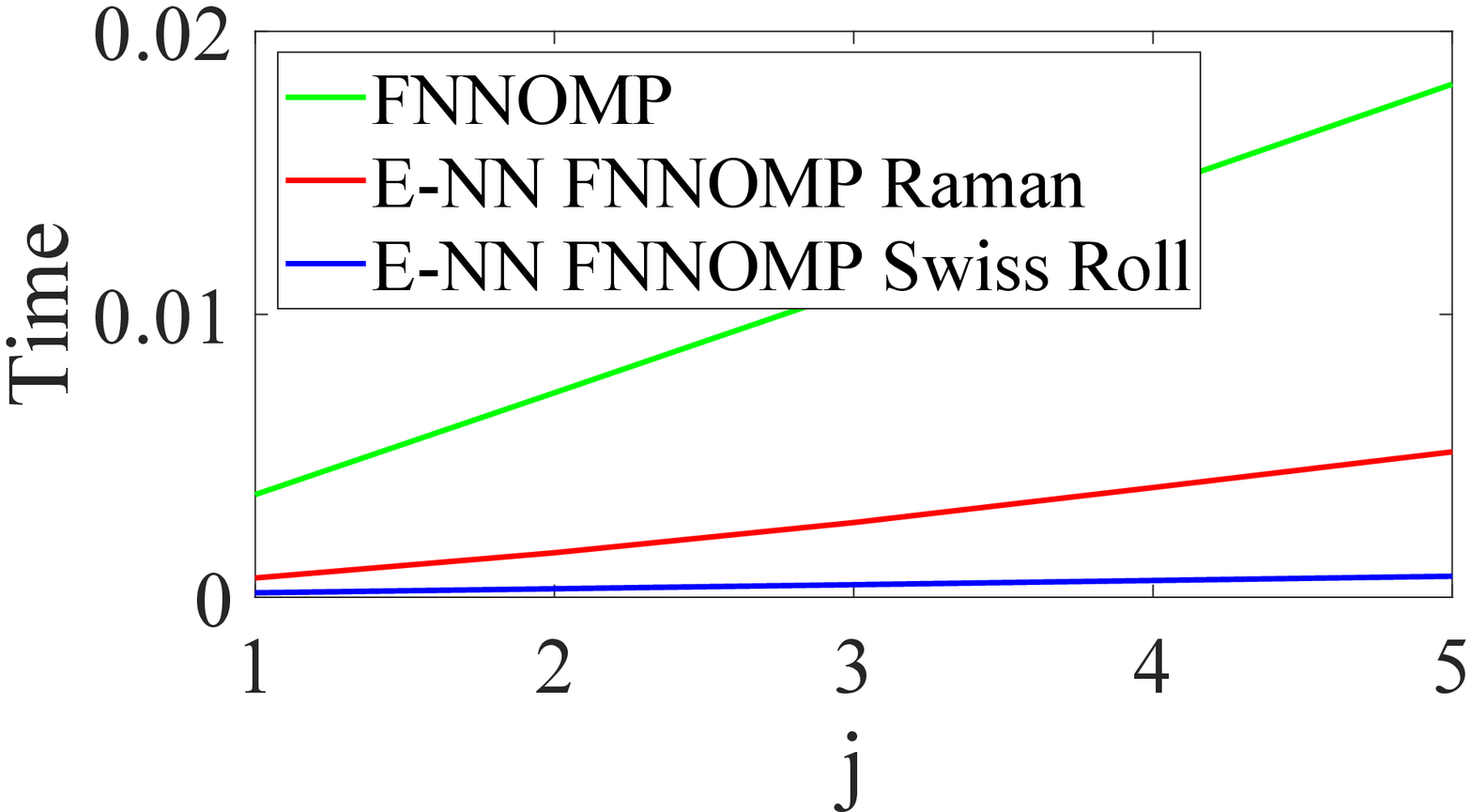}
   \caption{}
   \label{fig:Ng1} 
\end{subfigure}
\begin{subfigure}[b]{0.45\textwidth}
   \includegraphics[width=1\linewidth]{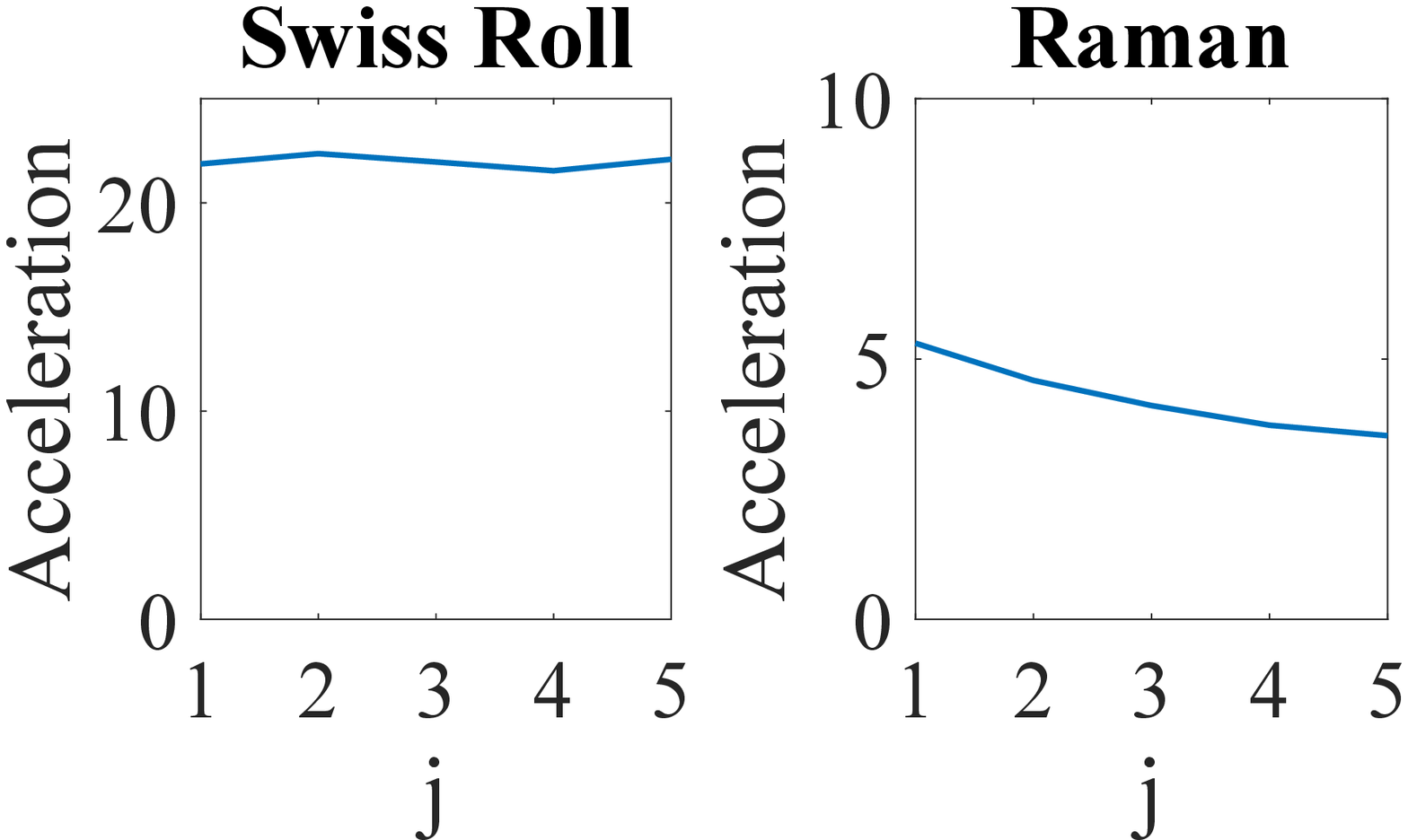}
   \caption{}
   \label{fig:Ng2}
\end{subfigure}\caption{Top of the figure: Elapsed time for each of the algorithms. Bottom: Acceleration over sparsity. Where \textrm{ Acceleration}$(j)= \frac{\textrm{Time FNNOMP}(j)}{\textrm{Time E-NN FNNOMP}(j)}$.}\label{fig:acceleration}
\end{figure}

\begin{figure}[!b]
\centering
\includegraphics[width=0.45\textwidth]{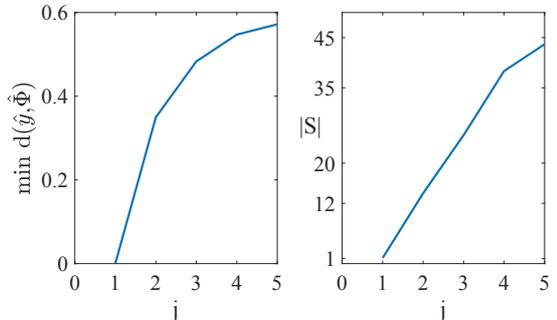}\vspace{-0.2cm} 
\caption{Average points in Step 4 of E\textendash NN  over sparsity.}\label{fig:cardinality}
\end{figure}

\section{Results}\label{sec:results}

In this section we evaluate  the  performance of the proposed algorithm with respect to FNNOMP. Based on the results introduced in figure \ref{fig:lower} we select the $Q$ obtained by the \textit{NuMax} algorithm as the linear operator that  projects offline the dictionary $\Phi$  and online the mixture  $y$ in $R^{K}$ while for the Swiss Roll we select the $Q$ obtained by PCA. We set $\delta=0.09$ for the \textit{Raman} library and $\delta=0$ for the \textit{Swiss Roll}. We then generate signal mixtures of  varying sparsity $j$ from the elements in $\Phi$.

The obtained  results demonstrated in figure \ref{fig:acceleration} show that E\textendash NN FNNOMP is generally faster than FNNOMP. The overall performance of the algorithm though decays over sparsity for the \textit{Raman} spectra. Given that the computational cost at steps 1 and 2 of E\textendash NN, the only parameter related to the complexity that may vary over $j$ is $|S|$. In order to obtain a better understanding regarding that issue we demonstrate the average number of points per iteration of the algorithm in figure \ref{fig:cardinality}. As can be seen from the results, the task of signal decomposition in the lower dimensional space becomes more difficult while sparsity increases. This is obviously not the case for the search in Swiss Roll. Essentially the acceleration factor remains constant. This happens because $\delta=0$ hence the update step of  E\textendash NN is unnecessary. This means that in practice we compare the implementation of FNNOMP into different domains. This phenomenon may occur when all of the points that lie in $R^M$ in reality they lie in the same subspace $R^{K}$. As it can be seen from the \textit{Raman} library though this is not something to be expected in a realistic setting.

\section{CONCLUSIONS}
We here presented  E\textendash NN which is a novel algorithm aiming to accelerate the NN sparse decomposition using a big library. The obtained results indicate that the E\textendash NN FNNOMP  outperforms FNNOMP. The current approach of E\textendash NN leverages the underlying sparsity of $\Phi$ via a linear embedding of $\Phi$  on $R^K$. However many datasets contain essential nonlinear structures that are invisible to linear techniques \cite{isomap}. For example, the \textit{Swiss Roll} dataset consists of 3D points that form a 2D manifold. PCA and \textit{NuMax} reveal the underlying linear subspace that our artificial dataset lives, but they cannot benefit from the underlying geometrical structure of this space. Exploring the nonlinear dimensionality reduction for acceleration of nonnegative sparse approximations has been left for the future work.

\begin{center}
Acknowledgment
\end{center}
This work was supported by the Engineering and Physical Sciences Research Council (EPSRC) Grant numbers EP/S000631/1 and EP/K014277/1 and the MOD University Defence Research Collaboration (UDRC) in Signal Processing.

\end{document}